# Magnetoelectric Response in Multiferroic $SrFe_{12}O_{19}$ Ceramics

*Guolong Tan[*], Yao Huang, Haohao Sheng,*

*State Key Laboratory of Advanced Technology for Materials Synthesis and Processing,*

*Wuhan University of Technology, Wuhan 430070, China*

**Abstract:**

We report here the realization of ferroelectricity, ferromagnetism and magnetocapacitance effect in $SrFe_{12}O_{19}$ ceramics at room temperature. The ceramics demonstrate a saturated polarization hysteresis loop, two I-V peaks and large anomaly of dielectric constant near Curie temperature. These evidences confirmed the ferroelectricity of $SrFe_{12}O_{19}$ ceramics after annealing in $O_2$ atmosphere. The remnant polarization of the $SrFe_{12}O_{19}$ ceramic is 103 $\mu C/cm^2$. The material also exhibits strong ferromagnetic characterization, the coercive field and remnant magnetic moment are 6192Oe and 35.8emu/g, respectively. Subsequent annealing $SrFe_{12}O_{19}$ ceramics in $O_2$ not only reveals its innate ferroelectricity but also improves the ferromagnetic properties through transforming $Fe^{2+}$ into $Fe^{3+}$. By applying a magnetic field, the capacitance demonstrates remarkable change along with B field, the maximum relative change in $\varepsilon$ *($\Delta\varepsilon(B)/\varepsilon(0)$)* is 1174%, which reflects a giant magnetocapacitance effect in $SrFe_{12}O_{19}$. These combined functional responses in $SrFe_{12}O_{19}$ ceramics opens substantial possibilities for applications in novel electric devices.

## 1. Introduction

Multiferroics is a class of functional materials that simultaneously exhibit ferroelectricity and ferromagnetism in a single structure[1-3]. They can demonstrate not only the magnetic or electric polarization but also the desired magnetoelectric (ME) coupling between the two orders leading to multifunctionalities such as electric field controlled magnetic data storage or vice versa [4]. This unique coupling feature has a tremendous impact on technology, with potential application for spintronic devices, solid-state transformers, high sensitivity magnetic field sensors, and actuators [5]. As part of the technological drive toward device miniaturization, considerable effort has been devoted to the combination of electronic and magnetic properties into one multifunctional material, i.e., a single device component that can perform more than one task [5-9]. The idea of having the two order parameters has stimulated a vast research of new multiferroic materials [10-16]. Some ferrites with hexagonal crystal structures, termed hexaferrites, have been found to show ME effects as magnetic field induced ferroelectrics at room temperature and low magnetic fields [17-19]. Although there are several ferrites that exhibit ME effects and remarkable changes in electric polarization in response to a magnetic field [20-23], the ME effect in these ferrites is small, their pure ferroelectric features (F-E loops) are still absent and the magnetism is weak [28]. For applications, however, it will be necessary to generate simultaneously ferroelectricity and ferromagnetism, together with giant ME effects at room temperature. Hence, it is a long standing challenge in the research of multiferroics to improve the operating temperature[24] and the ME sensitivity [19, 25].

[*] *Corresponding author, Tel: +86-27-87870271; fax: +86-27-87879468. Email address: gltan@whut.edu.cn*

M-type lead hexaferrite ($PbFe_{12}O_{19}$) has been reported to present simultaneously large spontaneous electric and magnetic polarization at room temperature [26, 27]. However, lead (Pb) is a kind of toxic element and $PbFe_{12}O_{19}$ is not an environment-friendly material. $SrFe_{12}O_{19}$, instead, is a lead-free M-type hexaferrite and environment-friendly. It has attracted a lot of attention because of its non-toxicity, excellent magnetic properties and wide application in various field such as magnetic recording and high-frequency devices [28, 29]. Recently, the dielectric and ferroelectric aspects of M-type hexaferrites have attracted some attentions[30-32]. S. P. Shen claimed that the M-type hexaferrites, such as $BaFe_{12}O_{19}$ single crystal, belong to quantum paraelectrics due to the electric dipole of a $FeO_5$ bipyramid[30, 31]. However, their XRD patterns of the $BaFe_{12}O_{19}$ and $SrFe_{12}O_{19}$ single crystals are completely different from the standard one of M-type hexaferrites, i.e., the strong diffraction peaks from {110}, {007} and {114} lattice planes are missing. Their crystals exhibited much higher symmetric structure than M-type hexaferrites. In addition, the single crystals were grown in a sealed furnace, where the oxygen concentration was very low. Therefore the single crystals would be in heavy oxygen deficiency, which could induce the formation of large numbers of oxygen vacancies and $Fe^{2+}$. Such crystals could cause large current leakage during the electronic measurement and would display a pseudo paraelectric phenomenon. The doubtful ferroelectric property of $SrFe_{12}O_{19}$ ceramics had also been reported several years ago in our previous study[33]; however, its ferroelectric hysteresis loops differ significantly from classic ferroelectric counterparts and resembled "bananas" due to the current leakage. Its ferroelectricity remains controversial and the banana-shaped P-E loops are not convincing evidence of the material's ferroelectricity [34]. The ME effect of $SrFe_{12}O_{19}$ has not been investigated yet. In this paper, we will optimize the fabrication process to remove the oxygen vacancies and transform $Fe^{2+}$ into $Fe^{3+}$ by subsequent annealing $SrFe_{12}O_{19}$ specimen in oxygen atmosphere so as to reduce the current leakage. We will then demonstrate the improved ferroelectric properties, dielectric anomaly near the Curie temperature, remarkable ME response, together with strong ferromagnetism in the M-type hexaferrite of $SrFe_{12}O_{19}$.

## 2. Experimental Procedure

We started with the preparation of nanocrystalline $SrFe_{12}O_{19}$ powders by polymer precursor procedure. Strontium acetate ($Sr(CH_3COO)_2 \cdot 3H_2O$) and ferric acetylacetonate ($C_{15}H_{21}FeO_6$) were used as starting material, which were bought from Afar Aesar. First of all, 0.2467g strontium acetate was dissolved in 15 mL glycerin to form a clear solution. The solution was distilled in a rotary evaporator at 120°C for 1 h to remove the water trapped in $Sr(CH_3COO)_2 \cdot 3H_2O$. The distilled solution was put in a 50 mL flask, which was moved into a glove box. 4.026 g of ferric acetylacetonate was weighed and dissolved in a mixture solution of 100 mL anhydrous ethanol and 70 mL acetone in a 250 mL three-neck flask inside the glove box. The solution was stirred at 70°C for 6 hours to ensure that ferric acetylacetonate was fully dissolved. Subsequently, the strontium and ferric precursor solutions were mixed together. Here, the molar ratio of strontium to iron was set to 1:9.5~10 to balance the Sr loss during the subsequent heat treatment process. Afterwards, 45 mL ammonia solution and 15 mL polyethylene glycol were poured into the above mixture solution. The dispersion solution was maintained at 70°C under stirring for 24 hours and then moved out of the glove box. The water and organic molecules were removed from the dispersion solution by centrifugation at 12000 rpm. The remaining colloid powders were calcined at 450°C for 1 hour. The powders were grinded in a agate mortar for 1 hour and then calcined again at 800°C for another hour to ensure total removal of organic components. Thus, pure $SrFe_{12}O_{19}$ powders in a single phase were obtained. 0.060 g of $SrFe_{12}O_{19}$ powders were weighed and pressed in a module into a pellet, which was then sintered at 1150°C for 1 hour into a solid ceramic specimen. The ceramic pellet was subsequently annealed in pure $O_2$ at 800°C for 3 hours. Then the ceramic pellet was turned over with upside down and the annealing process was repeated again for another 3 hours. After the furnace was cooling down to room temperature, the ceramic was heat-treated in pure $O_2$ once more at 700°C for another 3 hours. In this way, the oxygen vacancies could be removed and $Fe^{2+}$ would be fully transformed into $Fe^{3+}$, so as to greatly enhance the resistance of the ceramics and reduce the current leakage during the ferroelectric measurement. Phase identification of the $SrFe_{12}O_{19}$ powder and ceramic was performed by X-ray powder diffraction (XRD) with Cu–$K_\alpha$ radiation. Magnetization was measured using a Quantum Design physical property measurement system (PPMS). For P-E hysteresis loop measurement, both surfaces of the ceramic pellets were coated with silver paste as electrodes which was heat treated at 820°C for 15 min; the ferroelectric hysteresis loop was measured using a lab-constructed instrument, referred to as the ZT-IA ferroelectric measurement system. The temperature-dependent dielectric properties were measured by an LCR instrument (HP 8248A). The complex impedance spectrum was measured upon a electrochemical station (Chenghua) within the frequency range of 0.01 Hz ~ 1 MHz. The magnetocapacitance parameters of the $SrFe_{12}O_{19}$ pellet were measured using a Wayne Kerr 6500B LCR station by applying a variable magnetic field.

## 3. Results and Discussion

### 3.1 Structure and Electric Properties of $SrFe_{12}O_{19}$ compound

*Figure 1*a shows the X-Ray diffraction (XRD) pattern of our $SrFe_{12}O_{19}$ specimen, the underneath lines in red color are the standard diffraction spectrum of $SrFe_{12}O_{19}$ (PDF#33-1340). The single-phase $SrFe_{12}O_{19}$ powders has been sintered at 1150°C for 1h and subsequently annealed in $O_2$ for 9hs.

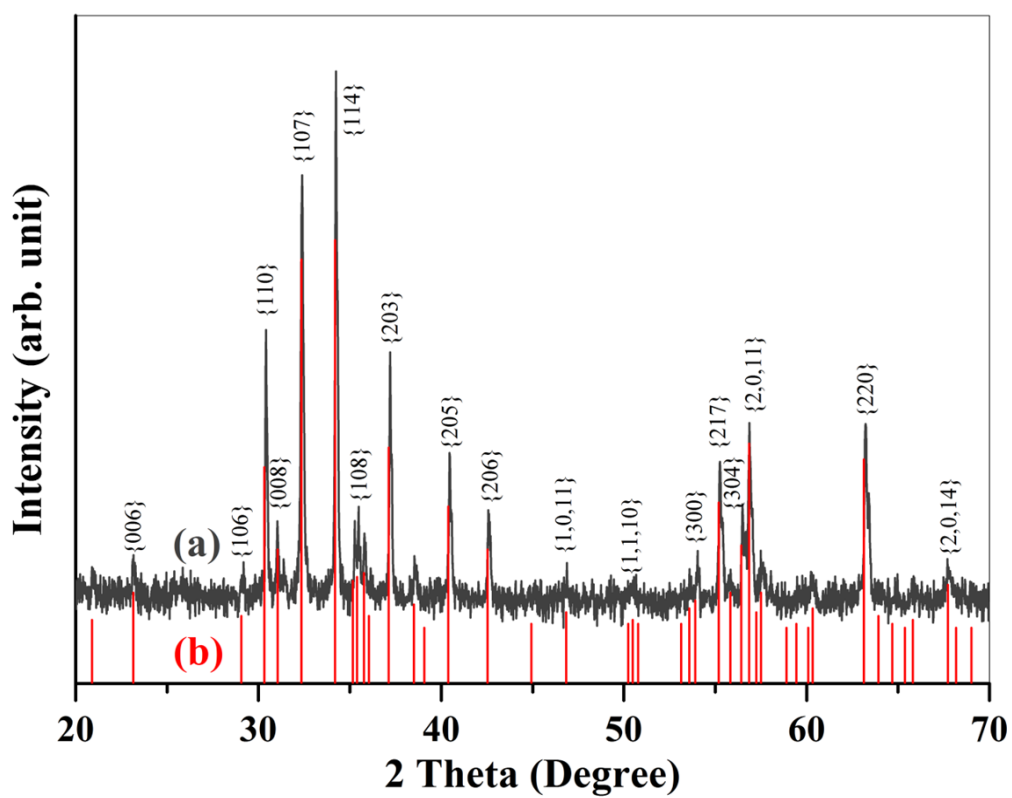

*Figure 1*: *(a), XRD pattern of $SrFe_{12}O_{19}$ with $O_2$ annealing process. (b) the standard diffraction pattern of the $SrFe_{12}O_{19}$ (PDF#33-1340) being marked by discrete red lines.*

It can be seen that all the diffractions peaks of the oxygen annealed specimen match well with the corresponding ones from the standard cards (*PDF#33-1340*), indicating the formation of pure $SrFe_{12}O_{19}$. No diffraction peaks from any second ferrite phases or impurity compounds have been indexed in the pattern, revealing the stability of the magnetoplumbite structure until 1150°C. This diffraction pattern is completely different from that of reported $SrFe_{12}O_{19}$ single crystals[30-32], which exhibit much higher symmetry and is lacking the strongest diffraction peaks form {110}, {007} and {114} lattice planes of typical M-type hexaferrites. Since the structure of our fabricated $SrFe_{12}O_{19}$ ceramics is different from those $SrFe_{12}O_{19}$ crystals reported in the literatures[30-32], the symmetry and electric properties should also differ significantly.

### 3.2 Dielectric Relaxation of $SrFe_{12}O_{19}$ ceramics

In order to check out if oxygen annealing process could remove the oxygen vacancies and transform $Fe^{2+}$ into $Fe^{3+}$, we measured the complex impedance spectrum of the $SrFe_{12}O_{19}$ ceramics with and without $O_2$ annealing by an electrochemical station. Figure 2(a) and (b) exhibits the module of complex impedance for the $SrFe_{12}O_{19}$ ceramics with and without oxygen heat-treatment process, respectively. The modules represents the magnitude of the impedance or resistance, which reflects concentration of oxygen vacancies and $Fe^{2+}$ in $SrFe_{12}O_{19}$ ceramics. The higher is the module, the lower is the concentration of these charge carriers.

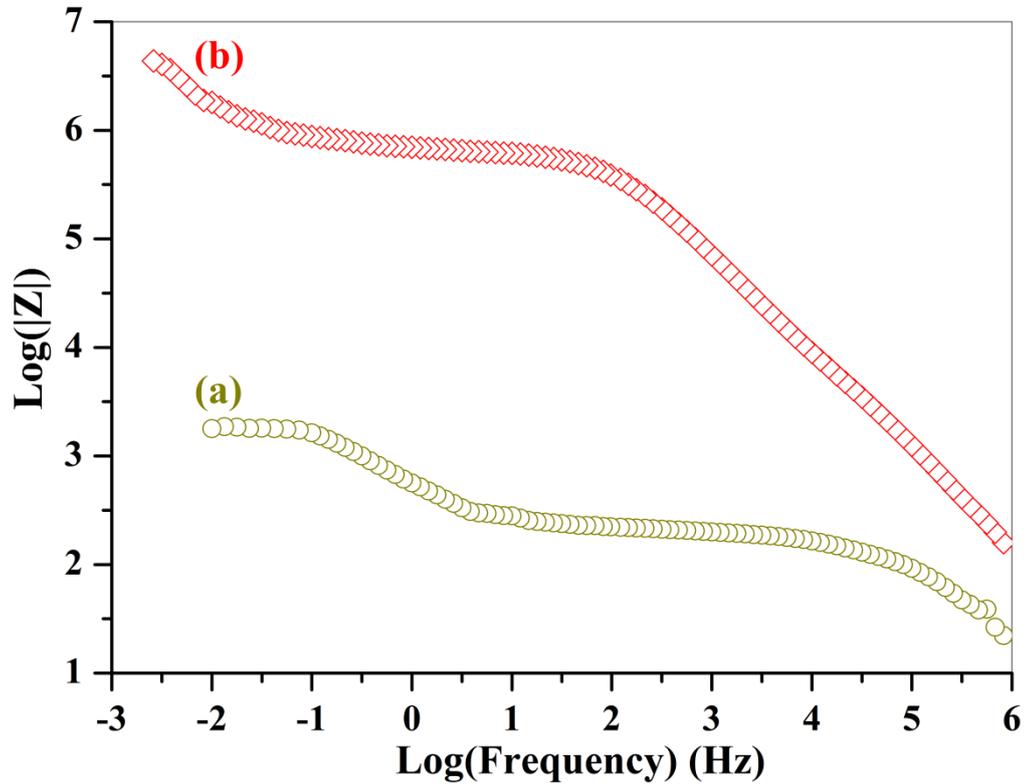

Figure 2：*Modules of the complex impedance for (a) the conventional $SrFe_{12}O_{19}$ ceramics being sintered at 1100°C in air only, (b) the sintered ceramic was subsequently annealed in pure oxygen atmosphere for 9hs..*

The impedance module or resistance of the specimen with $O_2$ heat-treatment is much higher than that of the ceramic without $O_2$ treatment within the whole frequency region. The electric resistance of $SrFe_{12}O_{19}$ ceramic with $O_2$ treatment is $7.9\times10^6$ Ω, while that of $SrFe_{12}O_{19}$ without $O_2$ annealing process is only $1.8\times10^3$ Ω at frequency of 0.01 Hz. The resistance (module) of $SrFe_{12}O_{19}$ ceramic was enhanced by a factor of 4389 after annealing in $O_2$. The great enhancement of the resistance reveals the drastic reduction of the concentration of the oxygen vacancies and full conversion of $Fe^{2+}$ into $Fe^{3+}$ ions, since the current leakage from oxygen vacancies and electronic hopping between $Fe^{2+}$ and $Fe^{3+}$ has been precluded.

Figure 3a represents the complex impedance spectrum of $SrFe_{12}O_{19}$ ceramic without subsequent $O_2$ annealing process. The spectrum is composed of a small Cole circle with a diameter of 215 and a large Cole one with a diameter of 1627. Each circle represents a circuit composed of a capacitor and a resistor which are connected in parallel. The two linked Cole circles could then be expressed as two such equivalent series connected circuits, as being shown in *Figure 4*. The small Cole circle contributes from the grain boundaries and the large one from the grains in $SrFe_{12}O_{19}$ ceramics. Figure 3(b) demonstrates a more complicated impedance spectrum for $SrFe_{12}O_{19}$ ceramics after $O_2$ annealing process. The equivalent circuit for the spectrum could also be expressed as two series linked circuits, each one is composed of a capacitor and a resistor being parallel connected (*Figure 4*). Each Cole circle corresponds to one individual circuit, one for grains and the other one for grain boundaries. Similarly, the spectrum is composed of a small Cole circle and a large half Cole circle, the diameter of the small one is estimated to be $9.0\times10^5$ and that of large one is $9.8\times10^6$. Obviously, the contribution of the impedance from both grain boundaries and grains in $SrFe_{12}O_{19}$ ceramics with subsequent $O_2$ heat-treatment have been greatly enhanced in comparison with that without $O_2$ annealing process. Both real and imaginary parts of the impedance have been promoted more than 1000 times after $O_2$ heat-treatment.

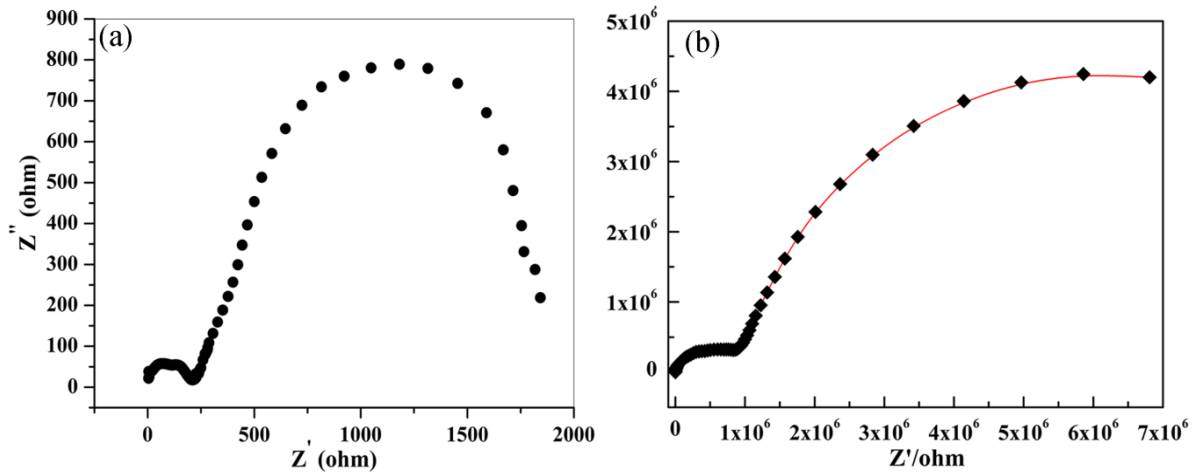

Figure 3: *Complex impedance spectrum of the $SrFe_{12}O_{19}$ ceramic within the frequency range of 0.01 Hz to 1 MHz, (a) for the ceramic being sintered at 1150 °C in air only; (b) for the sintered ceramic with subsequently annealing in $O_2$.*

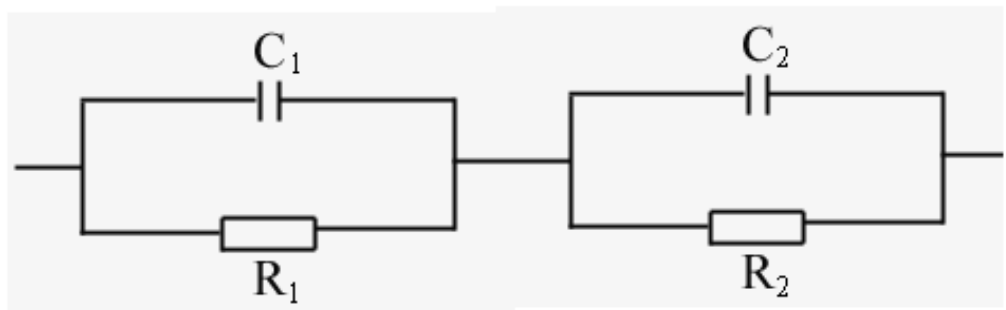

Figure 4: *The equivalent circuit for the complex impedance of $SrFe_{12}O_{19}$ ceramics, the scheme is composed of two series linked sub-circuits with a capacitor and a resistor parallelly connected.*

We then measured the dielectric relaxation behavior of $O_2$ heat-treated $SrFe_{12}O_{19}$ ceramics by a HP4284A LCR instrument. *Figure 5* shows the temperature-dependent dielectric constants of the specimen at different frequencies of 1kHz, 10kHz and 100 kHz. At 1kHz, there appear three peaks, locating at 174 °C, 368 °C and 490 °C (*Figure 5*a), the first two peaks are similar to that of $PbFe_{12}O_{19}$ corresponding to two kinds of phase-transitions[27]. Similarly, the first peak $T_d$ could be assigned to the ferroelectric to anti-ferroelectric phase transition, while the second one ($T_m$) to anti-ferroelectric to para-electric phase transition. The maximum dielectric constant at $T_d$ is 2621. The third peak is attributed from a complicated phase transition. Since the dielectric constant (ε') becomes negative when temperature is higher than 527 °C (*Figure 5*), the phase structure could then be assigned to a so-called "left hand materials (LHM)" whose dielectric constant is less than 0. Therefore the third peak ($T_l$=527 °C) is proposed to be the phase transition from para-electric phase to LHM. The first transition peak is sensitive to the frequency, the larger is the frequency, the higher is the transition temperature. When the frequency increases from 1 kHz to 10 kHz, the first transition peak shifts from 174 °C to 199 °C and the maximum dielectric constant drops from 2261 to 1394. Further increasing the frequency from 10 kHz to 100 kHz, this peak shifts to 239 °C and the maximum dielectric constant decreases from 1394 to 847. However, the second transition peaks didn't move accordingly, while the third peak ($T_l$) moves to the opposite direction, the higher is the frequency, the lower is the transition temperature.

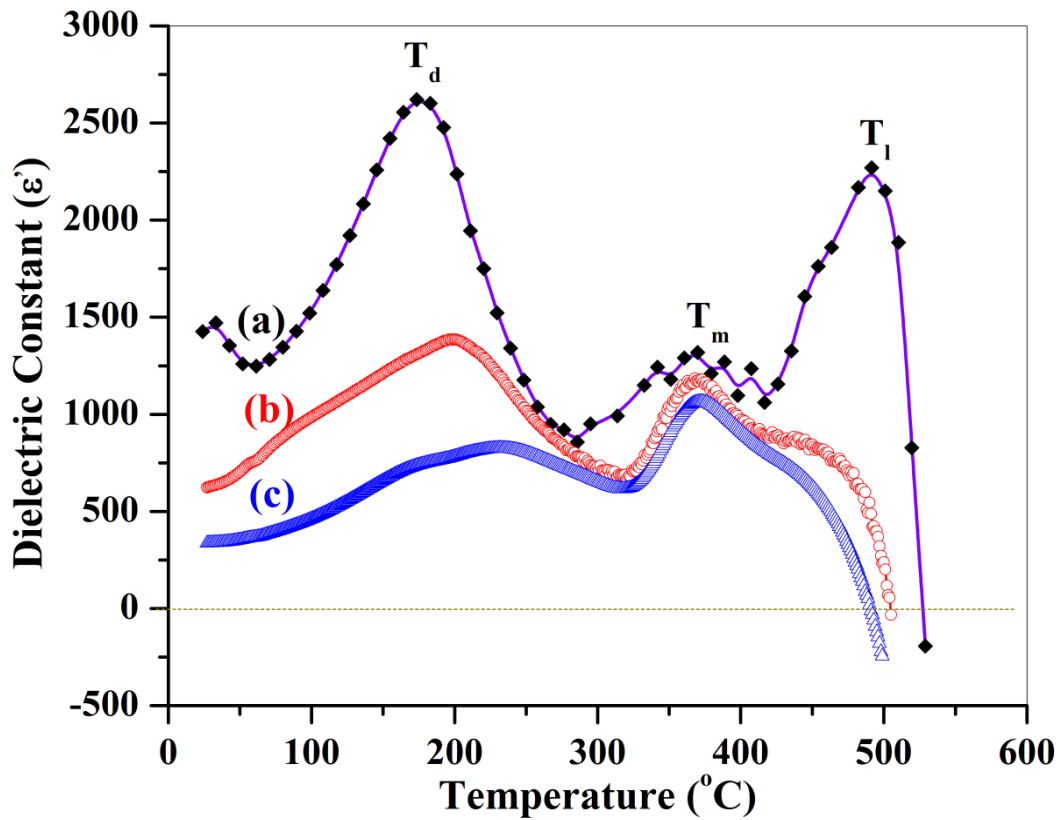

*Figure 5: Plot of dielectric constant as a function of temperature for SrFe$_{12}$O$_{19}$ ceramics at frequencies of (a) 1kHz, (b) 10kHz and (c) 100 kHz.*

The measurements at different frequencies show large shifts in the first peak positions, suggesting that $SrFe_{12}O_{19}$ is a relaxor ferroelectric compound with a diffuse phase transition. At 1kHz, the maximum ε-T peak (490°C) demonstrates a strong ferroelectric to antiferroelectric phase transition, associated with a broad ε (T) anomaly near the vicinity of the transition temperature. These kinds of dielectric anomalies at different frequencies provides additional evidence for the ferroelectricity of $SrFe_{12}O_{19}$.

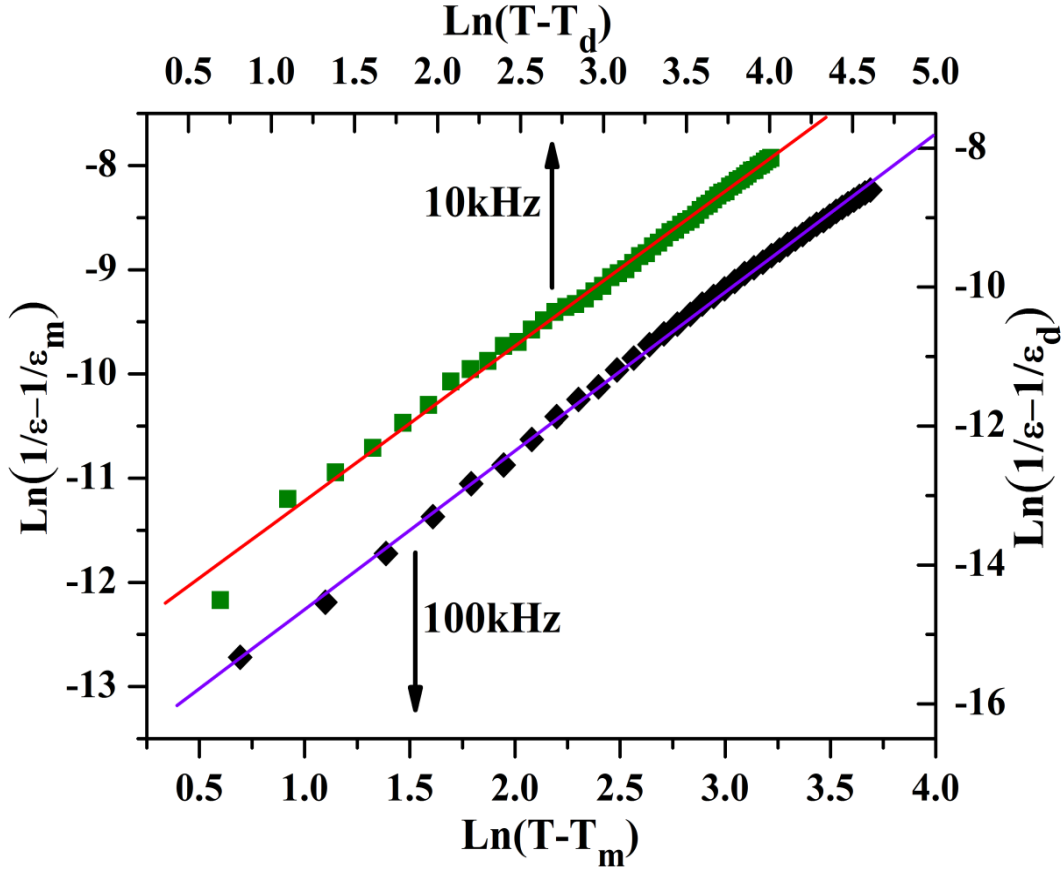

*Figure 6: Modified Curie-Weiss law calculation: (a) Logarithm of (1/ε-1/$ε_d$) as a function of logarithm of (T-$T_d$) at 10 kHz. and (b) logarithm of (1/ε-1/$ε_m$) as a function of logarithm of (T-$T_m$) at 100 kHz for the $SrFe_{12}O_{19}$ ceramic being sintered at 1150°C for 1 hour and subsequently annealed in $O_2$ for 9 hours.*

Upon the anomalies of the dielectric constants, we then made the calculation of the reciprocal dielectric constants as a function of temperature using modified Curie-Weiss law, being expressed as follows:

$$\frac{1}{\varepsilon} - \frac{1}{\varepsilon_m} = C(T - T_m)^\gamma$$

where γ is the critical exponent, representing the degree of diffuseness of the transition,

and *C* is a Curie–Weiss-like constant. $\varepsilon_m$ is the maximum dielectric constant at temperature of transition peak $T_m$. For a sharp transition, $\gamma = 1$, the materials are called normal ferroelectrics. Diffuse transitions lie in the range $1 < \gamma < 2$ [35], whereas at $\gamma = 2$ the materials correspond to a so-called "complete" diffuse phase. When $\gamma > 2$, the materials would take a diffuse phase transition from ferroelectrics to anti-ferroelectrics or antiferroelectrics to paraelectricity [27].

*Figure 6* shows plots of $Ln(1/\varepsilon - 1/\varepsilon_d)$ as a function of $Ln(T-T_d)$ at 10 kHz and $Ln(1/\varepsilon - 1/\varepsilon_m)$ as a function of $Ln(T-T_m)$ at 100 kHz for $SrFe_{12}O_{19}$ ceramic. Linear fitting to the experimental data using Curie-Weiss formula derives out the slope of the fitting lines, which were determined to be $\gamma =2.3$ and 2.2 at frequencies of 10 KHz and 100 kHz, respectively. The calculated lines following Curie-Weiss formula fit well with the experimental data points. The linear relationship between $Ln(1/\varepsilon - 1/\varepsilon_m)$ and $Ln(T-T_m)$ reveals that the temperature dependence of the dielectric constant obeys the Curie–Weiss law, suggesting the relaxor ferroelectric behavior of the $SrFe_{12}O_{19}$ ceramics.

## 3.2. Ferroelectric Polarization of $SrFe_{12}O_{19}$ ceramics

The ferroelectric loop of the $SrFe_{12}O_{19}$ ceramics without $O_2$ heat-treatment was looking like a "banana" [33], which had drawn lots of doubts on the validity of its ferroelectricity. Considering that the "banana" shaped electric loop could be induced by current leakage and the necessity of confirming the validity of its ferroelectricity, we then heat treated the $SrFe_{12}O_{19}$ ceramics in pure oxygen atmosphere for three times at 800 C for 9 hours, so as to greatly enhance its resistance. The great reduction of the concentration of charge carriers, such as oxygen vacancies and $Fe^{2+}$ by annealing the ceramics in oxygen, could dramatically reduce the current leakage and thus saturate the ferroelectric hysteresis loop of the $SrFe_{12}O_{19}$ ceramics.

During the ferroelectric measurement, the specimen was parallel-connected to a capacitor of 1 μF for compensation. *Figure 7*a shows a fully saturated ferroelectric hysteresis loop of the $SrFe_{12}O_{19}$ ceramic with $O_2$ annealing process. A drastic variation of the polarization appears in the vicinity of the specimen's coercive field at around 10 kV/m. Further increasing the applied field up to 25 kV/m, the polarization of the ceramic gradually approaches to saturation along with a concave arc line (*Figure 7*a).

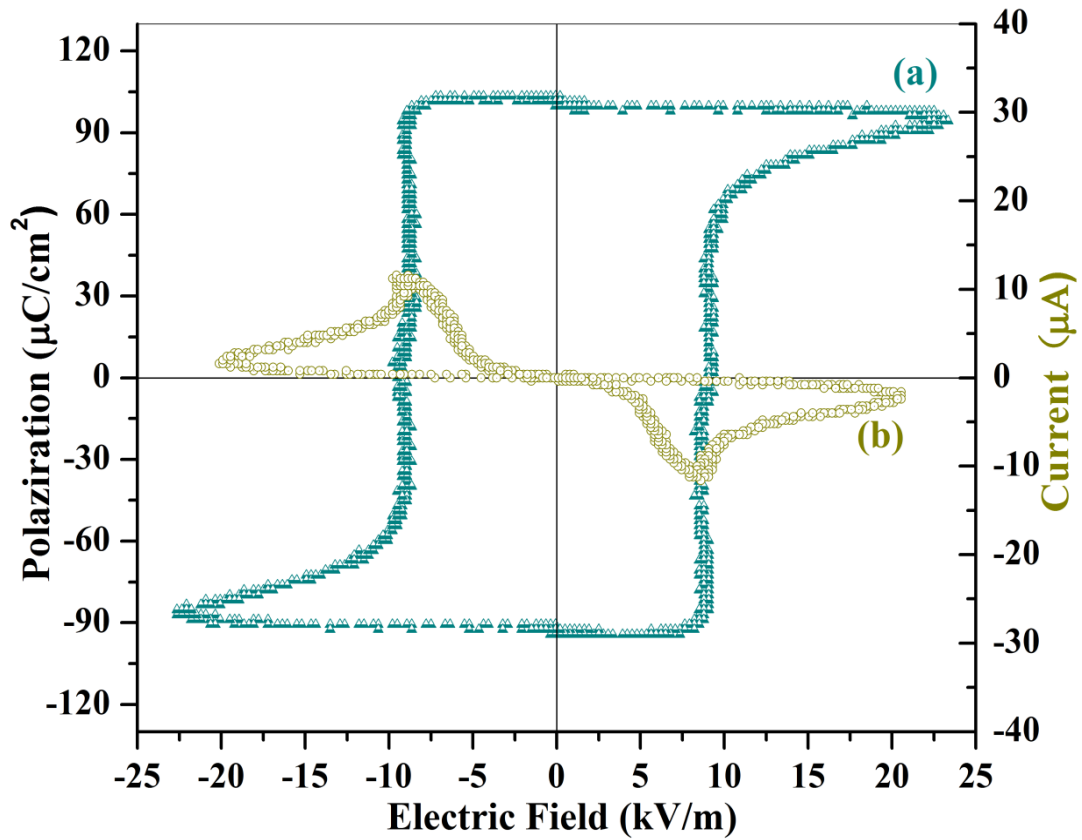

*Figure 7: (a) The saturated ferroelectric hysteresis loop, and (b) a plot of current as a function of voltage (I-V curve) of $SrFe_{12}O_{19}$ ceramic. The ceramic has been sintered at 1150°C for 1 hour and subsequently annealed at 800°C in pure oxygen for 9 hours. The measurement was made at a frequency of 33Hz and room temperature (300K).*

When the applied field decreases, the polarization remains at the value of saturation because most of the ceramic's domains still align themselves along the electric field's direction. The spontaneous polarization, which is equal to the saturation value of the electric displacement extrapolated to the zero-field strength, remains almost constant with external field variations. This result reveals that all the electric displacement dipoles have aligned themselves along the direction of the external field until the external field was less than the coercive field of the ceramics. When the applied field switched to the reversal direction, the spontaneous polarization demonstrated a hysteresis and changed the direction suddenly at the position of negative coercive field. The polarization voltage changes signs to be negative and approaches to the negative saturated value along with a reversal concave arc line (*Figure 7*a) within the field range of -10 kV/m to -25 kV/m. The remnant polarization in this classic hysteresis loop is estimated to be 103 µC/cm$^2$, which is around 8.3 times higher than that (15 µC/cm$^2$) of SrFe$_{12}$O$_{19}$ ceramics without subsequent heat treatment in O$_2$ [33]. Therefore subsequent annealing SrFe$_{12}$O$_{19}$ ceramics in oxygen not only saturated the hysteresis loop, but also greatly improved the remnant polarization value through reducing the current leakage, which results from the removal of oxygen vacancies and the transformation of Fe$^{2+}$ to Fe$^{3+}$ [26, 27]. Similar ferroelectric hysteresis loops being measured on different SrFe$_{12}$O$_{19}$ ceramic specimens are supplied in the Supplementary Materials, so as to confirm the reliability and repetitiveness of ferroelectric data

The last evidence for the validity of ferroelectricity for SrFe$_{12}$O$_{19}$ ceramics would be attributed from the appearance of two current peaks along with the polarization switching (I-V curve). When the ferroelectric polarization is switching, the screening surface charges flow from one electrode to the other one and create momentarily a sudden change of current. In the current versus voltage plot, this will result in two peaks with reversal directions as being shown in *Figure 7* (b). The two nonlinear I-V peaks show very clearly the switching phenomenon of the polarization and does not present any linear current component (or current leakage component). The two I-V peaks are similar to that of typical ferroelectric compounds (Pb(Zr$_{0.4}$Ti$_{0.6}$)O$_3$ and LiNbO$_3$) [36] and could convince us that the F-E hysteresis loop indeed origins from the ferroelectric polarization instead of current leakage. The origin of the ferroelectricity of SrFe$_{12}$O$_{19}$ ceramics is similar to that of PbFe$_{12}$O$_{19}$ and has been discussed in detail in our previous literatures [26, 27], since both compounds share the same crystal structure. The off-center shift of the Fe$^{3+}$ ions and the displacement of O$^{2-}$ ions from its original corner positions in the FeO$_6$ octahedron are supposed to be the origin of polarization in SrFe$_{12}$O$_{19}$ [26, 27]. The saturated ferroelectric hysteresis loop, two peaks in the I-V curve, the giant anomalies of the dielectric constant in the vicinity of the transition temperatures (T$_m$ and T$_d$) as well as the comply of the reciprocal dielectric constant with modified Curie Weiss law provide us with enough evidences to prove the intrinsic ferroelectricity of SrFe$_{12}$O$_{19}$ ceramics.

This result is quite different from those reported M-type hexaferrtie single crystals, which were claimed to be a new family of magnetic quantum paraelectrics and retained paraelectric symmetry down to zero temperature[30-32]. Actually those crystals showed different crystal structure with higher symmetry and were grown in a sealed furnace without subsequent heat-treatment in oxygen atmosphere. There would be high concentration of oxygen vacancies and $Fe^{2+}$ ions inside the crystals. These kinds of carrier charges would reduce the electric resistance of the crystals and induce large current leakage during the electric measurement. As such no saturated polarization hysteresis loop could be observed in these crystals. Both different crystal structure and low concentration of carrier charges make our ceramic specimens differ significantly from those crystals in ferroelectric and dielectric properties.

### 3.3 Magnetic Properties of $SrFe_{12}O_{19}$ Compound

For magnetic measurement, the $SrFe_{12}O_{19}$ powders were prepared with the same heat-treatment history as that of above ceramics. The magnetic measurement was made upon $SrFe_{12}O_{19}$ powders by the Physical Property Measurement System (PPMS) at room temperature. *Figure 8* exhibits the ferromagnetic hysteresis loops of $SrFe_{12}O_{19}$ powders with and without $O_2$ heat-treatment. It can be seen that magnetic properties of $SrFe_{12}O_{19}$ have been greatly improved by annealing the powders in oxygen atmosphere. The coercive fields of the $SrFe_{12}O_{19}$ powders with $O_2$ treatment reaches as high as 6192 Oe, while that of the same powders without $O_2$ treatment is 4111 Oe. The coercive field has been promoted 2081 Oe through $O_2$ heat-treatment. The remnant magnetic moment has also been enhanced from 33.5 emu/g to 35.8 emu/g after annealing the $SrFe_{12}O_{19}$ powders in oxygen atmosphere. Such promotion of magnetic polarization was also observed in M-type lead hexaferrite ($PbFe_{12}O_{19}$) after annealing in $O_2$ atmosphere [27].

The heat-treatment in oxygen atmosphere transformed $Fe^{2+}$ into $Fe^{3+}$，which provides one unpaired electron spin for magnetic polarization of $SrFe_{12}O_{19}$. $Fe^{2+}$, however, has no unpaired electron spins and thus would not induce magnetic polarization. A certain content of $Fe^{2+}$ existing in $SrFe_{12}O_{19}$ which was sintered in a oxygen deficient atmosphere, such as a sealed air furnace, would reduce its ability for magnetic polarization and degrade its magnetic properties. Therefore heat-treatment of $SrFe_{12}O_{19}$ in oxygen would not only improve the ferroelectric polarization performance but also enhance the ferromagnetic properties through transforming $Fe^{2+}$ into $Fe^{3+}$. The large hysteresis loop reflects the strong magnetic feature of $SrFe_{12}O_{19}$. The above combined results demonstrate the simultaneous occurrence of large ferroelectricity and strong ferromagnetism in the single $SrFe_{12}O_{19}$ compound at room temperature. It allows us to expect a new generation of electronic devices being made of such a practicable multiferroic candidate, in which large ferroelectricity and strong ferromagnetism coexist

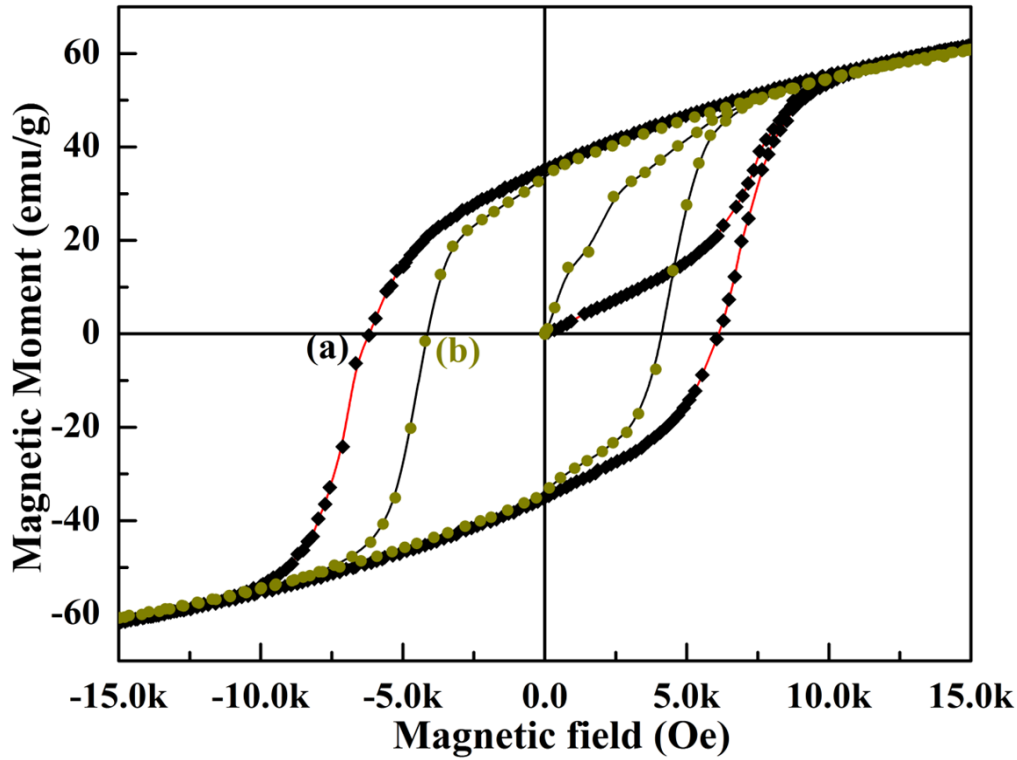

*Figure 8*: *Magnetic hysteresis loop of $SrFe_{12}O_{19}$ (a) being sintered at 1150°C for 1 h and subsequently annealed in $O_2$ for 9 hs, (b) without heat-treatment in $O_2$.*

.

### 3.4 **Magnetocapacitance Effect of $SrFe_{12}O_{19}$ Ceramics**

Previous studies for certain rare-earth manganites [6, 37] suggest that materials having long wavelength magnetic structures often exhibit a strong interplay between magnetic ordering and ferroelectricity, which makes the capacitance of the manganites [38] and Y-type hexaferrites exhibit great response to the B field [39]. In order to check out if the M-type strontium hexaferrite ($SrFe_{12}O_{19}$) could also generate such coupling response upon an external field, we set up a simple system for the ME coupling measurement, which was performed by measuring the capacitance as a function of the magnetic field (B). The SrFe$_{12}$O$_{19}$ ceramic was coated with silver electrodes on both sides and then placed in a space between two electromagnets. Upon the application of the magnetic field, the Wayne Kerr 6500B LCR Precision impedance analyzer, which was linked with the electrodes on both surfaces of the ceramic, would output the variable capacitance with the external magnetic field B. The B-field-dependent relative magnetic permeability was calculated using a defined formula, which can be expressed as follows[40]:

$$\mu_r = \frac{Z(T)-Z(0)}{if\mu_0 h \ln\frac{c}{b}} + 1 = [\frac{Z''(T)-Z''(0)}{f\mu_0 h \ln\frac{c}{b}} + 1] - i[\frac{Z'(T)-Z'(0)}{f\mu_0 h \ln\frac{c}{b}}] \quad (1)$$

where Z(T) is the complex impedance when magnetic field B=T, Z(0) the impedance for B=0; $\mu_0$ is the vacuum permeability, h is the space between two magnets, c and b are the inner and outer radius of the ring magnets. *Figure 9* displays the dependence of the relative magnetic permeability ($\mu_r$) on B and the change in ε (or magnetocapacitance) along with B field. It can be seen from *Figure 9*a that $\mu_r$ increases in a stepwise fashion, which is attributed to the evolution in magnetic structures. Five successive magnetoelectric phases could be thus mapped out: the first terrace (0<B<55mT) for modified helix, the ramp (55mT<B<150mT) for intermediate I, the second terrace (140mT<B<480mT) for intermediate II and III. The collinear ferrimagnetic phase corresponds to the B range with a rather large slope in $\mu_r$ (480mT<B<883mT). These magnetic phases has been delimited by yellow dot lines (*Figure 9*a).

*Figure 9*b displays great response of the capacity (or ε) to the applied magnetic field (B). The dramatic variation of dielectric constant with B field is similar to that of M-type BaFe$_{12-x}$Sc$_x$O$_{19}$ [37] and Y-type hexaferrite Ba$_{0.5}$Sr$_{1.5}$Zn$_2$(Fe$_{1-x}$Al$_x$)$_{12}$O$_{22}$ [18], and is associated with the magnetic phase transition along with B field. Within helix and intermediate I phases in range of 0~150 mT, the dielectric constant displays a downward slope line with small declining value less than 200. However, when B comes into the Intermediate II phase, the dielectric constant increases rapidly from -243 to 2650. The great change in ε is performed as two remarkable peak structures centered in the middle of Intermediate II (254 mT) and III (363 mT) phases respectively (*Figure 9*b). The maximum dielectric constants at the centers of the two peaks are determined to be 2650

and 2340, respectively. The dielectric constants show rapid drops at the magnetic boundary between Intermediate II and III (254 mT<B<300 mT) and that between Intermediate III and collinear phases (300 mT<B<48 mT). The valley bottom ( =66) between the two peaks locates at 300 mT, which could assigned to the borderline between intermediate II and III magnetic phases.

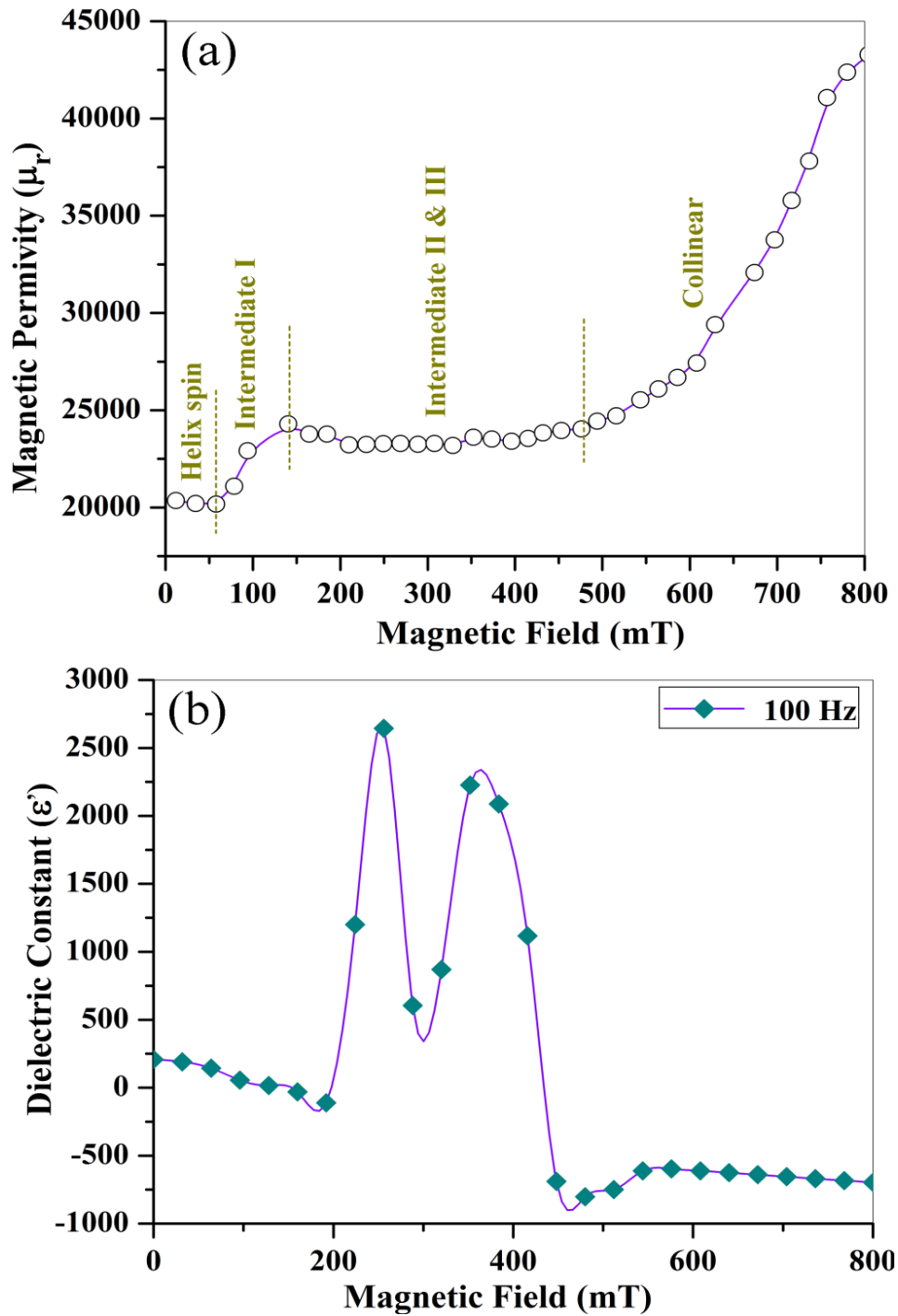

*Figure 9: Magnetocapacitance effect: (a) Magnetic permeability ($\mu_r$) and (b) variable*

*dielectric constant as a function of B field for the $O_2$ treated $SrFe_{12}O_{19}$ ceramics at a frequency of 100Hz and room temperature (300K).*

When B comes across into ferrimagnetic collinear phase (B>480), the dielectric constant remains almost as a flat line ( $\varepsilon$ ~-730) with little fluctuation, whose amplitude is less than 90. The maximum relative change in $\varepsilon$ $(\Delta\varepsilon(B)/\varepsilon(0)=[\varepsilon(B)-\varepsilon(0)]/\varepsilon(0)$, where B=254mT) is 1174%, which reflects a giant magnetocapacitance effect of $SrFe_{12}O_{19}$ ceramics. Thus, the hexaferrite having long-wavelength magnetic structures exhibits remarkable ME responses at low B field and room temperature, which opens substantial possibilities for applications of ME systems.

In brief, large ferroelectricity and strong ferromagnetism are naturally merged together in $SrFe_{12}O_{19}$, due to the coexistence of the off-centered $FeO_6$ octahedron in its unit cell and electron spins in partially filled 3d orbits of the $Fe^{3+}$ ions. Thus, the mutually exclusive electric and magnetic orders are integrated in $SrFe_{12}O_{19}$. Therefore, large ferroelectricity, strong ferromagnetism and giant ME coupling effect are all synchronously realized in one single phase of $SrFe_{12}O_{19}$ at room temperature (300K).

### 3.5 Magnetoelectric Coupling Effects of $SrFe_{12}O_{19}$ ceramics

In order to check out if $SrFe_{12}O_{19}$ could generate coupling charge upon an external magnetic field. We set up a simple system for the ME coupling measurement, which was performed by measuring the output coupling voltage (V) as a function of the magnetic field (B). The $SrFe_{12}O_{19}$ ceramic was coated with silver electrodes on both sides and then placed in a space between two electromagnets. Upon the application of the magnetic field, the microvoltmeter, which was linked with the electrodes on both surfaces of the sample, would output the variable coupling voltage with the external magnetic field B. *Figure 10*s shows such ME coupling voltages as a function of B at room temperature. By applying a low magnetic field from 0 to 50 mT, the coupling voltage varies from 0.04 mV to 0.1 mV, the coupling voltage changes very little within this range. With further increase of B from 50 mT to 200 mT, the coupling voltage shows a rapid enhancement from 0.1 mV to 3 mV, which then shows a slight decrease down to 2.2 mV when the magnetic field extends from 200 mT to 337 mT (*Figure 10*s). The variance ratio of the coupling voltage upon magnetic field B is around 7400%, which reflects a strong ME coupling effect. Then the coupling voltage exhibits a little fluctuation with a very small amplitude within the B range of 337 mT to 620 mT. Afterwards, with further increase of B up to 850 mT, the coupling voltage drops rapidly from 2.6 mV down to 0.3 mV (*Figure 10*).

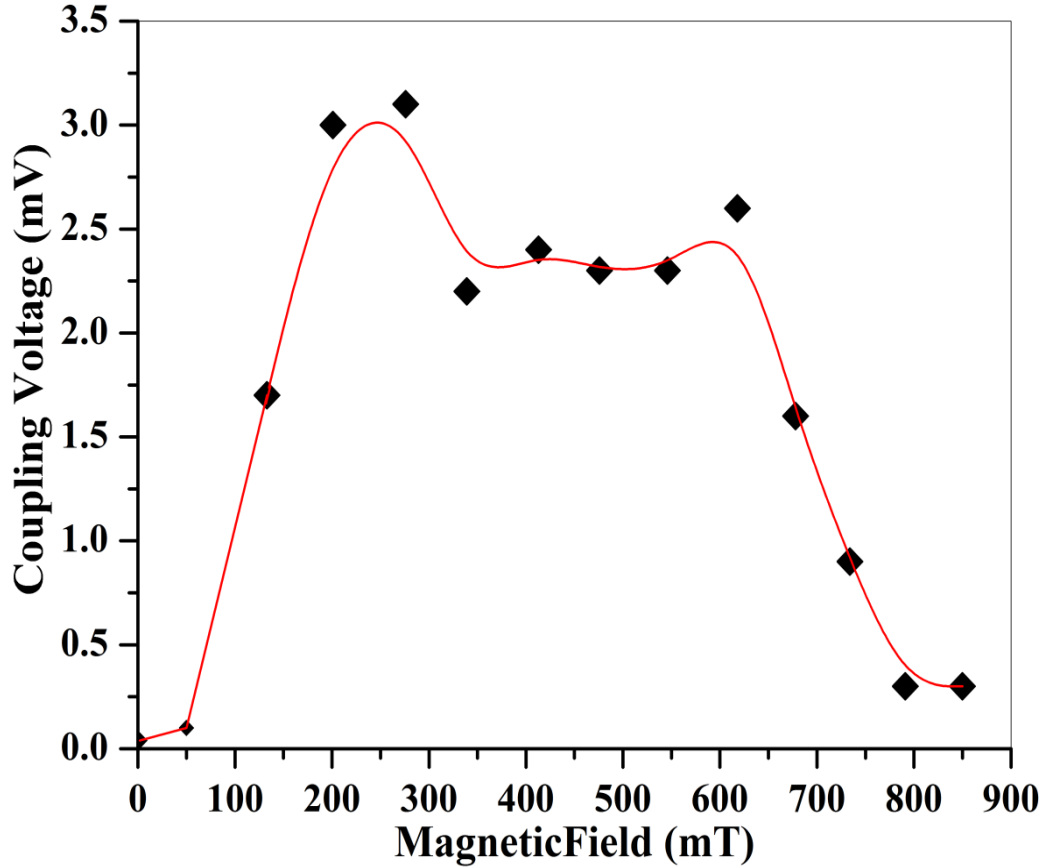

*Figure 10: Plot of magnetoelectric coupling voltage as a function of applied magnetic field for SrFe$_{12}$O$_{19}$ ceramics.*

### 3.6 Polarization Hysteresis Loops from Different Specimens

*In order to verify the ferroelectricity of SrFe$_{12}$O$_{19}$ ceramics, we measured the P-E hysteresis loops for several different specimens. These hysteresis loops are all fully saturated, the loop shapes are similar to the standard polarization hysteresis loops of such classic ferroelectric compounds as BiFeO$_3$, BasTiO$_3$, and PZT, indicating the intrinsic ferroelectricity of SrFe$_{12}$O$_{19}$ ceramics. These results are displayed in Figure 11 and Figure 12.*

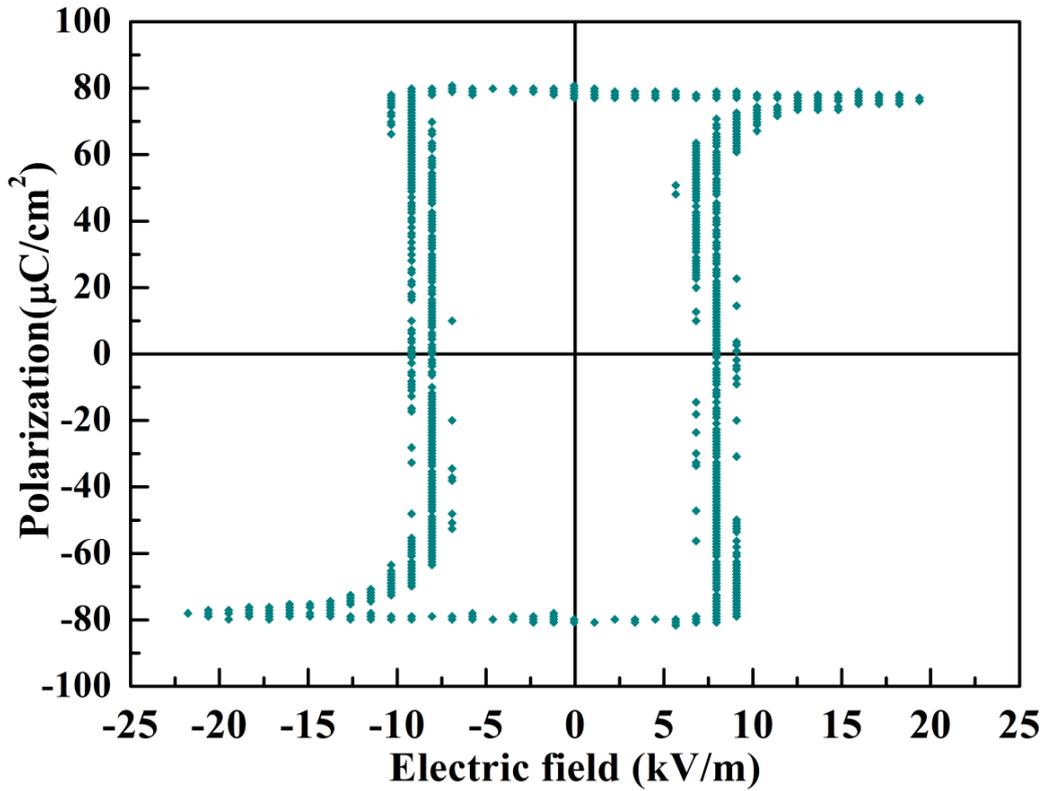

*Figure 11: Ferroelectric hysteresis loop being measured on the second SrFe$_{12}$O$_{19}$ ceramic specimen with subsequent heat-treatment in O$_2$.*

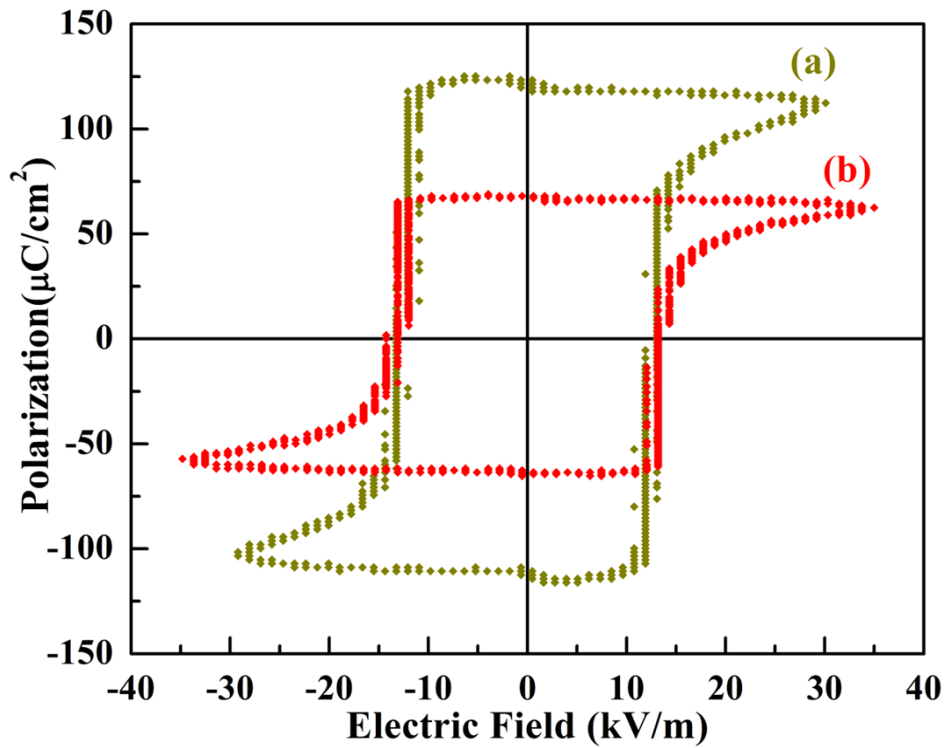

*Figure 12: Ferroelectric hysteresis loop being measured on the third SrFe$_{12}$O$_{19}$ ceramic specimen with the same annealing process in oxygen for different applied field.*

## 4. Conclusion

In summary, our work directly demonstrates the coexistence of large ferroelectricity and strong ferromagnetism in M-type strontium hexaferrite ($SrFe_{12}O_{19}$) at room-temperature (300K). we not only merge together the electric and magnetic orders, but also realize the giant magnetocapacitance effect in $SrFe_{12}O_{19}$ at room temperature. The $SrFe_{12}O_{19}$ ceramic displays a classical polarization hysteresis loop with full saturation, two particular nonlinear I–V peaks, and dielectric anomalies near the Curie temperature, all of which verify its intrinsic ferroelectricity. Subsequent annealing the $SrFe_{12}O_{19}$ ceramic in oxygen atmosphere greatly enhance its electric resistance through removal of oxygen vacancies and transformation of $Fe^{2+}$ into $Fe^{3+}$, leading to the full saturation of the F-E loop. The remnant polarization of the $SrFe_{12}O_{19}$ ceramics is 103 $\mu C/cm^2$. Large magnetic hysteresis loop was also observed in $SrFe_{12}O_{19}$ due to its strong ferromagnetism. Furthermore, the capacitance (or dielectric constant) exhibits dramatic variation along with B field. Two remarkable peak structures of $\varepsilon$ appeared at the centers of Intermediate II and III phases, respectively. The maximum relative change in $\varepsilon$ is 1174%, which reflects a giant magnetocapacitance effect of $SrFe_{12}O_{19}$ ceramics.

## 5. Acknowledgements:

The authors acknowledge the financial support from the National Natural Science Foundation of China under the contract of 51272201; Hubei Natural Science Foundation under the contract No. 2014CFB166; as well as Open fund of State Key Laboratory of Advanced Technology for Materials Synthesis and Processing (Wuhan University of Technology) under the contract No. 2016-KF-15.